\documentclass[%
reprint,
superscriptaddress,
showpacs,preprintnumbers,
 amsmath,amssymb,
 aps,
]{revtex4-1}

\usepackage{graphicx}% Include figure files
\usepackage{float}

\newcommand{\myfig}[4][ht]{
\begin{figure}[#1]
\centering
\includegraphics[#2]{#3}
\caption{#4\label{#3}}
\end{figure}
}

\graphicspath{{images/}}

\usepackage{dcolumn}% Align table columns on decimal point
\usepackage{bm}% bold math
\begin{document}
\title{Quasiballistic heat removal from small sources studied from first principles}
\author{Bjorn Vermeersch}%
 \email{Email: bjorn.vermeersch@cea.fr}%
\affiliation{CEA, LITEN, 17 Rue des Martyrs, 38054 Grenoble, France}%
\affiliation{Universit\'e Grenoble Alpes, France}%
\author{Natalio Mingo}%
\affiliation{CEA, LITEN, 17 Rue des Martyrs, 38054 Grenoble, France}%
\affiliation{Universit\'e Grenoble Alpes, France}%
\vspace{2ex}
\date{\today}
\begin{abstract}
Heat sources whose characteristic dimension $R$ is comparable to phonon mean free paths display thermal resistances that exceed conventional diffusive predictions. This has direct implications to (opto)electronics thermal management and phonon spectroscopy. Theoretical analyses have so far limited themselves to particular experimental configurations. Here, we build upon the multidimensional Boltzmann transport equation (BTE) to derive universal expressions for the apparent conductivity suppression $S(R) = \kappa_{\text{eff}}(R)/\kappa_{\text{bulk}}$ experienced by radially symmetric 2D and 3D sources. In striking analogy to cross-plane heat conduction in thin films, a distinct quasiballistic regime emerges between ballistic ($\kappa_{\text{eff}} \sim R$) and diffusive ($\kappa_{\text{eff}} \simeq \kappa_{\text{bulk}}$) asymptotes that displays a logarithmic dependence $\kappa_{\text{eff}} \sim \ln(R)$ in single crystals and fractional power dependence $\kappa_{\text{eff}} \sim R^{2-\alpha}$ in alloys (with $\alpha$ the L\'evy superdiffusion exponent). Analytical solutions and Monte Carlo simulations for spherical and circular heat sources in Si, GaAs, Si$_{0.99}$Ge$_{0.01}$ and Si$_{0.82}$Ge$_{0.18}$, all carried out from first principles, confirm the predicted generic tendencies. Contrary to the thin film case, common approximations like kinetic theory estimates $\kappa_{\text{eff}} \simeq \sum S_{\omega}^{\text{grey}} \, \kappa_{\omega}$ and modified Fourier temperature curves perform relatively poorly. Up to threefold deviations from the BTE solutions for sub-100$\,$nm sources underline the need for rigorous treatment of multidimensional nondiffusive transport.
\end{abstract}

\maketitle

%\tableofcontents

\section{Introduction}
Semiconductor thermal transport at short length scales has attracted growing interest over the past decade \cite{cahillreview,MFPspectroscopy,collins,minnich-TTG3D,siemens,hoogeboom,nanograting,TTG-fulldetails,TTG-GaAs,TTG-membranes,minnich-spotsize,minnich-VRMC,wilson,minnich-BTE3D,maznev,minnich-BTE1D,JAP-multidimensional}. In addition to its immediate relevance to thermal management of (sub)microscale electronics \cite{cahillreview}, this field of study has opened up pathways to phonon spectroscopy techniques \cite{MFPspectroscopy,collins,minnich-TTG3D} which help illuminate the fundamentals of heat conduction in nonmetallic materials.
\par
The self-heating of sources whose physical dimension is comparable to characteristic phonon mean free paths is known to exceed conventional (Fourier diffusion) predictions. Notable experiments, listed alongside the characteristic source dimension they impose, include coherent x-ray thermal probing of stripline arrays \cite{siemens,hoogeboom,nanograting} (line width $50\,\text{nm} \lesssim W \lesssim 100\,\mu\text{m}$), transient thermal grating (TTG) \cite{TTG-fulldetails,TTG-GaAs,TTG-membranes} (grating wavelength $2\,\mu\text{m} \lesssim \lambda \lesssim 25\,\mu\text{m}$), and time-domain thermoreflectance (TDTR) \cite{minnich-spotsize,minnich-VRMC,wilson} (laser $1/e^2$ beam radius $1\,\mu\text{m} \lesssim R \lesssim 30\,\mu\text{m}$).
\par
Steady advances have been made in the theoretical understanding of these experimental observations, along with how they can be exploited to reconstruct (portions of) the phonon spectrum. Most studies are conducted through explorations of the Boltzmann transport equation (BTE) under the relaxation time approximation (RTA). Siemens and coworkers \cite{siemens} used a minimal heated region model to explain the excess ballistic interface resistivity they measured between narrow stripline arrays and a saphire substrate. Hua and Minnich \cite{minnich-BTE3D} employed the analytic Green's function of the 3D isotropic BTE to compute the quasistatic heating profile for a $3 \omega$-like setting (single heater line on a semi-infinite substrate). Maznev \cite{maznev}, Collins \cite{collins} and Minnich \cite{minnich-BTE1D,minnich-TTG3D} and their respective co-workers investigated the effective conductivity suppression observed in TTG measurements. Ding and coworkers \cite{minnich-VRMC} conducted Monte Carlo simulations of a TDTR experiment to extract the radial conductivity suppression function. Vermeersch \cite{JAP-multidimensional} introduced a stochastic framework, capturing all essential physics of the multidimensional nondiffusive heat flow in compact form, that enables direct quasiballistic interpretation of raw TDTR and TTG measurement signals. 
\par
Most of the prior studies cited above yielded individual insights for a particular experimental setting. In this work, we complement the existing body of knowledge by revealing the generic prototypical tendencies in the effective conductivity reduction around small sources regardless of the precise details of their spatial dissipation signature. In section \ref{sec:theory}, we build upon the fundamental steady state solution of the multidimensional BTE to derive universal expressions for the isotropic conductivity suppression function for 3D and 2D source dissipations. These closed forms reveal an intermediate quasiballistic regime for which the effective conductivity in single crystals and alloys displays a respectively logarithmic and fractional power dependence on characteristic source dimension. Section \ref{sec:results} illustrates our findings with concrete examples through first-principles semi-analytic solutions and Monte Carlo simulations for spherical and circular sources in Si, GaAs, and two SiGe alloys. In section \ref{sec:discussion} we discuss important limitations of kinetic theory and modified Fourier approaches in describing the quasiballistic heat constriction. We additionally point out ultraviolet divergences within the actual BTE. A brief summary in section \ref{sec:conclusions} conludes the paper.
\section{Theory}\label{sec:theory}
\subsection{Methodology}\label{sec:theory:methodology}
We will analyse thermal conduction in and around small heat sources through the RTA-BTE under the additional assumption that the thermal dynamics can be considered as isotropic. This simplifies the mathematics considerably while still maintaining a high level of physical accuracy \cite{minnich-BTE3D,JAP-multidimensional,APL-thinfilms}, especially for cubic crystals. Under the stated conditions and at time scales exceeding phonon relaxation times, the single pulse response for the deviational thermal energy density $P \equiv C_{v} \, \Delta T$ (with $C_v$ the volumetric heat capacity) is known \cite{minnich-BTE3D,JAP-multidimensional} to take the following form in Fourier-Laplace domain ($\vec{r} \leftrightarrow \vec{\xi} , t \leftrightarrow s)$
\begin{equation}
P(\vec{\xi},s) = C_v \, \Delta T(\vec{\xi},s) = \frac{1}{s + \psi(\| \vec{\xi} \|)}
\end{equation}
The propagator function $\psi(\| \vec{\xi} \|)$ can be derived directly from wavevector-resolved phonon heat capacities $C$, relaxation times $\tau$ and mean free paths $\Lambda = \|\vec{v}\| \tau$ as \cite{JAP-multidimensional}
\begin{equation}
\psi(\| \vec{\xi} \|) = \sum \limits_{k} \frac{C_k \, \| \vec{\xi} \|^2 \Lambda_{x,k}^2}{\tau_k (1 + \| \vec{\xi} \|^2 \Lambda_{x,k}^2)} \, \biggr / \sum \limits_{k} \frac{C_k}{1 + \| \vec{\xi} \|^2 \Lambda_{x,k}^2} \label{psi_phonons}
\end{equation}
where the subscript $_x$ denotes cartesian projection.
\par
In this work we will focus on the steady state response ($s = 0$) induced by radially symmetric heat sources having characteristic dimension $R$ and volumetric/surface power dissipation $p(\| \vec{r} \|)$. The temperature rise $\Delta T(\vec{r} = \vec{0})$ at the center of the source is mathematically speaking easiest to obtain, but is rarely accessible experimentally and also suffers from the largest stochastic uncertainties in Monte Carlo simulations. As a more representative metric, we will instead determine a spatially averaged source temperature
\begin{equation}
\left< \Delta T \right>(R) = \iiint \limits_{\text{source}} w(\| \vec{r} \|) \, \Delta T (\| \vec{r} \|) \, \mathrm{d} \vec{r}
\end{equation}
in which $w$ denotes a properly normalised weight function i.e. $\iiint w(\| \vec{r} \|) \, \mathrm{d} \vec{r} = 1$.
\par
Solutions in diffusive regime $\psi(\| \vec{\xi} \|) = D \| \xi \|^2$ always scale inversely proportional to the bulk conductivity i.e. $\left< \Delta T \right>_{\text{Fourier}} \sim 1/\kappa$. Quasiballistic solutions $\left< \Delta T \right>_{\text{BTE}}$ for small sources are found to exceed their Fourier counterparts, which may be interpreted as if the source experiences a lower-than-nominal effective thermal conductivity. The associated suppression function is defined by
\begin{equation}
S(R) \equiv \frac{\kappa_{\text{eff}}(R)}{\kappa_{\text{bulk}}} = \frac{\left< \Delta T \right>_{\text{Fourier}}(R)}{\left< \Delta T \right>_{\text{BTE}}(R)} \label{genericS}
\end{equation}
\subsection{Universal formula for suppression function $S(R)$}\label{sec:theory:S}
\subsubsection{Volumetric dissipation (3D source)}\label{sec:theory:S3d}
The real-space temperature rise versus radial coordinate follows from 3D isotropic Fourier inversion \cite{fouriertransforms}
\begin{equation}
\Delta T(r) = \frac{1}{2 \pi^2 C_v} \, \int \limits_{0}^{\infty} \zeta^2 \, j_0(\zeta r) \, \frac{p(\zeta,R)}{\psi(\zeta)} \, \mathrm{d}\zeta
\end{equation}
where $j_0(\zeta r) \equiv \mathrm{sinc}(\zeta r)$ and $\zeta \equiv \sqrt{\xi_x^2 + \xi_y^2 + \xi_z^2}$. Spatial averaging yields
\begin{eqnarray}
\left< \Delta T(r) \right> & = & 4\pi \, \int \limits_{0}^{\infty} r^2 \, w(r,R) \, \Delta T(r) \, \mathrm{d}r \nonumber \\
& = & \frac{1}{2 \pi^2 C_v} \, \int \limits_{0}^{\infty} \zeta^2 \, w(\zeta,R) \, \frac{p(\zeta,R)}{\psi(\zeta)} \, \mathrm{d}\zeta
\end{eqnarray}
Dimensional arguments show that both $p(\zeta,R)$ and $w(\zeta,R)$ are always expressable as unitary functions of normalised variable $u = \zeta R$. Applying (\ref{genericS}) produces
\begin{equation}
S(R) = \frac{R^2 \, \int_{0}^{\infty} p(u) \, w(u) \, \mathrm{d}u}{\int_{0}^{\infty} p(u) \, w(u) \, \frac{D u^2}{\psi(\zeta = u/R)} \, \mathrm{d}u} \label{S_3Dsource}
\end{equation}
This result is valid for any radially symmetric source dissipation $p$ and weight function $w$. One particular example of prototypical interest is worth mentioning:
\begin{multline}
\text{spherical source with radius $R$ :} \\
\frac{p(u)}{Q_{\text{tot}}} = w(u) = \frac{3}{u^3} \, \left( \sin u - u \, \cos u \right) \label{sphericalsource}
\end{multline}
in which $Q_{\text{tot}}$ denotes the total source power.
\subsubsection{Surface dissipation (2D source)}\label{sec:theory:S2d}
The real-space temperature response inside the source plane, which under the isotropic assumption can be taken as $z=0$ without loss of generality, follows from inverse 1D Fourier transform with respect to the $z$ coordinate and subsequent Hankel inversion:
\begin{multline}
\underline{G_0(h)} = P(h,z=0) = \frac{1}{\pi} \, \int \limits_{0}^{\infty} \frac{\mathrm{d}\xi_z}{\psi(\zeta = \sqrt{h^2 + \xi_z^2})} \\
\Delta T(\rho,z=0) = \frac{1}{2\pi C_v} \, \int \limits_{0}^{\infty} h \, J_0(h \rho) \, p(h,R) \, \underline{G_0(h)} \, \mathrm{d}h
\end{multline}
where $J_0$ is the zeroth-order Bessel function of the first kind and $h \equiv \sqrt{\xi_x^2 + \xi_y^2}$. It is useful to note here that for Fourier diffusion one obtains $G_0(h) = 1/2hD$.
\par
After spatial averaging
\begin{equation}
\left< \Delta T(\rho,z=0) \right> = 2\pi \, \int \limits_{0}^{\infty} \rho \, w(\rho,R) \, \Delta T(\rho,z=0) \, \mathrm{d}\rho
\end{equation}
and change of integration variable $v \equiv hR$ we find
\begin{equation}
S(R) = \frac{\pi R \, \int_{0}^{\infty} p(v) \, w(v) \, \mathrm{d}v}{2D \, \int_{0}^{\infty} \mathrm{d}v \, v \, p(v) \, w(v) \, \int \limits_{0}^{\infty} \frac{\mathrm{d}\xi_z}{\psi \left( \zeta = \sqrt{(v/R)^2 + \xi_z^2} \right)}} \label{S_2Dsource}
\end{equation}
Once again this formula holds for any radially symmetric source dissipation $p$ and weight function $w$. Two examples of practical interest include
\begin{itemize}
\item{circular source with radius $R$ :}
\begin{equation}
\frac{p(v)}{Q_{\text{tot}}} = w(v) = \frac{2 J_1(v)}{v} \label{circularsource}
\end{equation}
\item{Gaussian pump/probe with $1/e^2$ radius $R$ :}
\begin{equation}
\frac{p(v)}{Q_{\text{tot}}} = w(v) = \exp \left( - v^2/2 \right) \label{gaussiansource}
\end{equation}
\end{itemize}
\subsection{Generic trends}\label{sec:theory:trends}
Building upon the previous section and published prototypical forms \cite{JAP-multidimensional} for the propagator function $\psi(\zeta)$ in single crystals and alloys, we can deduce several key features of the quasiballistic suppression of heat removal from small sources. It should be noted that the simple parametric forms for $S(R)$ derived below are not targeted at accurate quantitative fitting of first-principles predictions or experimental observations but merely aim to elucidate generic qualitative trends. The derived behaviour is then validated later in the paper by first-principles solutions and simulations.
\subsubsection{Single crystals}\label{sec:theory:trends:singlecrystal}
The full ballistic-diffusive transition in non-alloy materials can be described by the following propagator
\begin{equation}
\psi(\zeta) = \frac{D \zeta^2}{\ln(2)} \, \ln \left[ 1 + \left(1 + r_{\text{CF}}^{b} \, \zeta^{b} \right)^{-1/b} \right]
\end{equation}
This form is suggested by generic tendencies observed in first-principles computations and has been verified to yield accurate fits (with $b = 1.06$) to transient thermal grating measurements on bulk GaAs \cite{JAP-multidimensional}. The transition length scale $r_{\text{CF}}$ is on the order of dominant phonon mean free paths (typically about 1 micron). Assuming $b=1$ here for the sake of mathematical convenience, the resulting function can be approximated with good accuracy (deviations remain within a few percent) by
\pagebreak[4]
\begin{equation}
\psi(\zeta) \simeq \frac{D \zeta^2}{1 + r_0 \zeta} \qquad r_0 = 0.70 \, r_{\text{CF}} \label{psi_singlecrystal}
\end{equation}
Applying (\ref{S_3Dsource}) we obtain for a 3D source
\begin{equation}
\kappa_{\text{eff}} \simeq \frac{\kappa}{1 + A \cdot (r_0/R)} \quad , \quad A = \frac{\int_{0}^{\infty} u \, p(u) \, w(u) \, \mathrm{d}u}{\int_{0}^{\infty} p(u) \, w(u) \, \mathrm{d}u} \label{singlecrystal3D}
\end{equation}
A spherical source (\ref{sphericalsource}) yields $A = 15/4\pi \simeq 1.19$. Very small sources ($R \ll r_0$) operate ballistically ($\kappa_{\text{eff}} \sim R$) while large sources ($R \gg r_0$) achieve their nominal diffusive performance $\kappa_{\text{eff}} \simeq \kappa$ as expected. The intermediate quasiballistic portion of the curve appears quasilinear when plotted versus a logarithmic abcissa and thus approximately follows a $\kappa_{\text{eff}} \sim \ln(R)$ tendency.
\par
Attempting to evaluate the 2D source suppression function (\ref{S_2Dsource}) for the propagator (\ref{psi_singlecrystal}) leads to a divergent $\xi_z$ integral, which in turn produces the nonphysical result that $S \rightarrow 0 \leftrightarrow \left< \Delta T \right> \rightarrow \infty$ for all source dimensions $R$. This behaviour is akin to an `ultraviolet catastrophe', as will be discussed in more detail in section \ref{sec:discussion:uvdivergence}. These mathematical anomalies notwithstanding, we expect the physical tendencies for 3D sources to remain preserved for 2D counterparts, as Monte Carlo simulations for circular sources indeed confirm later in the paper. Additional experimental support is found in coherent x-ray results for Ni lines on a sapphire substrate \cite{siemens} that accurately follow the functional form (\ref{singlecrystal3D}) with $A \,  r_0 = 120\,$nm. 
\subsubsection{Alloys}\label{sec:theory:trends:alloy}
Prior theoretical \cite{PRB-levy1} as well as experimental \cite{PRB-levy2,PRB-rms,nanoletters,JAP-multidimensional} evidence has shown that semiconductor alloys exhibit L\'evy dynamics in their quasiballistic thermal regime:
\begin{equation}
\psi(\zeta) \simeq D_{\alpha} \zeta^{\alpha}  \quad , \quad D_{\alpha} = \frac{D}{r_{\text{LF}}^{2-\alpha}} \label{psi_alloys}
\end{equation}
where $1 < \alpha < 2$ is the characteristic superdiffusion exponent and $D_{\alpha}$ is referred to as the fractional diffusivity (unit m$^{\alpha}$/s). Conventional Fourier diffusion gives way to L\'evy dynamics around characteristic length scale $r_{\text{LF}}$ which is again found to be on the order of dominant phonon mean free paths. We note that eventually the superdiffusive regime in turn transitions to fully ballistic behaviour \cite{PRB-levy1}. Alloys will therefore exhibit the same ballistic ($\kappa_{\text{eff}} \sim R$) and diffusive ($\kappa_{\text{eff}} \simeq \kappa$) asymptotes also found in single crystals; here we focus our attention on the distinct L\'evy regime in between.
\par
Inserting (\ref{psi_alloys}) into (\ref{S_3Dsource}) and (\ref{S_2Dsource}) we respectively obtain
\begin{itemize}
\item{volumetric dissipation (3D source)}
\begin{equation}
\kappa_{\text{eff}} \simeq B_3 \, \left( \frac{R}{r_{\text{LF}}} \right)^{2-\alpha} \kappa \quad , \quad B_3 = \frac{\int_{0}^{\infty} p(u) \, w(u) \, \mathrm{d}u}{\int_{0}^{\infty} u^{2-\alpha} \, p(u) \, w(u) \, \mathrm{d}u} \label{alloy3D}
\end{equation}
\item{surface dissipation (2D source)}
\begin{multline}
\kappa_{\text{eff}} \simeq \frac{\sqrt{\pi} \,\, \Gamma \left( \frac{\alpha}{2}\right)}{\Gamma \left( \frac{\alpha-1}{2}\right)} \,\, B_2 \, \left( \frac{R}{r_{\text{LF}}} \right)^{2-\alpha} \, \kappa \quad , \\
B_2 = \frac{\int_{0}^{\infty} p(v) \, w(v) \, \mathrm{d}v}{\int_{0}^{\infty} v^{2-\alpha} \, p(v) \, w(v) \, \mathrm{d}v} \label{alloy2D}
\end{multline}
\end{itemize}
Notice that both source types display the same fractional dependence $\kappa_{\text{eff}} \sim R^{2-\alpha}$. The prefactors are only weakly dependent on $\alpha$ and for a typical alloy ($\alpha \simeq 5/3$) amount to $B_3 \simeq 1.02$ and $B_2 \simeq 1.05$ for the spherical source and $B_2 \simeq 1.31$ for the Gaussian case.
\section{First-principles results}\label{sec:results}
To illustrate the impeded heat removal from small sources with some concrete examples, we carry out first-principles analysis at room temperature $T_{\text{ref}} = 300\,$K for two single crystals (Si and GaAs) and two alloy compounds (Si$_{0.99}$Ge$_{0.01}$ and Si$_{0.82}$Ge$_{0.18}$). We performed DFT computations for each of the materials with \textsc{vasp} \cite{vasp} and obtained interatomic force constants through \textsc{phonopy} \cite{phonopy} as described in earlier work \cite{PRB-levy1}, and then used \textsc{almabte} \cite{almabte} to obtain phonon dispersions and scattering rates over a $24 \times 24 \times 24$ wavevector grid.
\subsection{Semi-analytic solution for spherical source}\label{sec:results:analytic}
Inserting the parameter-free phonon properties into Eq. (\ref{psi_phonons}) produces the propagator function $\psi(\zeta)$, which we then use to evaluate the suppression factor for spherical sources per Eqs. (\ref{S_3Dsource}) and (\ref{sphericalsource}).
\par
We achieve high numerical accuracy by isolating the oscillatory portion of the integrand and processing it fully analytically. Concretely, utilising the fact that
\begin{multline}
\int \frac{1}{u^5} \, \left( \sin u - u \cos u \right)^2 \, \mathrm{d}u \\ = -\frac{1}{4u^4} \left[ u^2 + \sin u \left( \sin u - 2u \cos u \right) \right] \equiv F(u)
\end{multline}
we obtain
\begin{equation}
S(R) \simeq \left( \frac{15D}{\pi R^2} \, \sum_n \frac{\bar{u}_n \, \left[ F(u_{n+1}) - F(u_n) \right]}{\psi(\zeta = \bar{u}_n / R)} \right)^{-1} \label{Ssphere_numerical}
\end{equation}
which is then evaluated over a logarithmically spaced $u$ grid with nodes $u_n$ and center points $\bar{u}_n = (u_n + u_{n+1})/2$. To validate this numerical scheme, we evaluated (\ref{Ssphere_numerical}) for the prototypical $\psi$ functions given earlier in Eqs. (\ref{psi_singlecrystal}) and (\ref{psi_alloys}); the results reproduced the respective analytic suppression functions (\ref{singlecrystal3D}) and (\ref{alloy3D}) within 0.003\%.
\subsection{Deviational Monte Carlo simulations}\label{sec:results:vrmc}
Variance-reduced Monte Carlo (VRMC) simulations help validate the above tendencies for spherical sources, and become a virtual necessity for circular sources in light of mathematical divergences in their analytic solutions.
\par
VRMC provides efficient numerical solutions of the linearised RTA-BTE by determining the deviational energy distributions (i.e. those in excess/deficit of the analytic Bose-Einstein reference distribution) from the stochastic trajectories of totally independent deviational particles \cite{peraud,almabte}. Each of the $N$ particles represents a fixed fraction $Q_{\text{tot}}/N$ of the dissipated source power. The contribution of each trajectory segment to the thermal field is easily evaluated through multidimensional extension of the 1D algorithm implemented within \textsc{almabte}.
\par
A small complication arises: lest computational particles be tracked indefinitely, VRMC simulations are unable to reproduce true `infinite' medium solutions. Instead, a spatial cutoff is introduced beyond which the particle can be considered out of view so that the simulation can proceed launching the next one. In practice, we terminate the simulation domain by an isothermal reservoir at $T_{\text{ref}}$. The reservoir absorbs all incoming particles and acts as an ideal heat sink that enforces $\Delta T = 0$ at the boundary. The finite extent of the geometry and isothermal boundary condition is rigorously accounted for in the calculation of the Fourier solutions needed to determine the suppression function, as detailed shortly.
\par
We performed our simulations with $N = 10^5$ particles, affording excellent stochastic repeatability (the normalised standard deviation $\sigma_S/S$ among 10 independent simulations remained between $\pm 0.2\%$ and $\pm0.4\%$ for all considered source dimensions). The power density of the source is set such that $(\Delta T)_{\text{max}} \simeq 1\,\text{K}$, thereby assuring that the simulation remains well within the linear regime.
\subsubsection{Spherical source}\label{sec:results:vrmc3d}
The simulation geometry is sketched in Fig. \ref{fig1_sphericalgeometry}.%
\myfig[!htb]{width=0.3\textwidth}{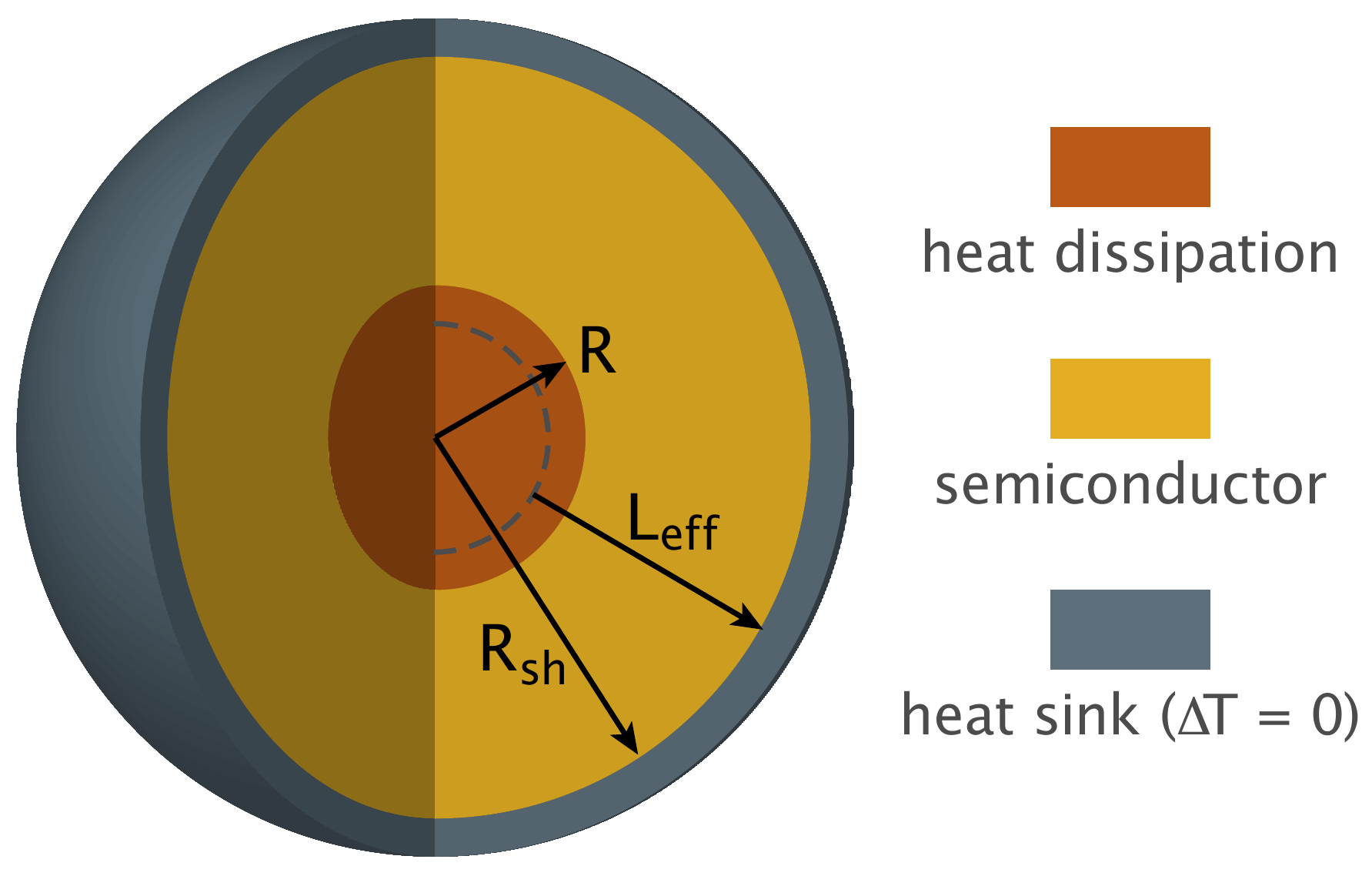}{Simulation geometry for spherical heat source.}
\par
The heat sink sits a distance $L_{\text{eff}}$ away from the average particle launch radius $3R/4$, i.e. the shell radius is set to $R_{\text{sh}} = \frac{3}{4} \, R + L_{\text{eff}}$. The diffusion equation for the considered geometry is easily solved analytically:
\pagebreak[4]
\begin{eqnarray}
\Delta T_{\text{Fourier}}(r) & = & \frac{p_v R^2}{2 \kappa} \left[ 1 - \frac{1}{3} \left( \frac{r}{R} \right)^2 - \frac{2}{3} \frac{R}{R_{\text{sh}}} \right] \,\, 0 \leq r \leq R \nonumber \\
\Delta T_{\text{Fourier}}(r) & = & \frac{p_v R^3}{3 \kappa} \left[ \frac{1}{r} - \frac{1}{R_{\text{sh}}} \right] \quad R \leq r \leq R_{\text{sh}} \label{Fourier_sphericalfield}
\end{eqnarray}
where $p_v = 3 Q_{\text{tot}}/4\pi R^3$ denotes the volumetric power density of the source. Spatial averaging yields
\begin{equation}
\left< \Delta T_{\text{Fourier}} \right> =  \frac{p_v R^2}{15 \kappa} \cdot \left( 6 - \frac{5R}{R_{\text{sh}}} \right)
\end{equation}
Comparing this diffusive solution to the average source temperature observed in the VRMC simulations produces suppression curves per Eq. (\ref{genericS}) for various $R$ and $L_{\text{eff}}$ combinations (Fig. \ref{fig2_spherical}).
\myfig[!htb]{width=0.48\textwidth}{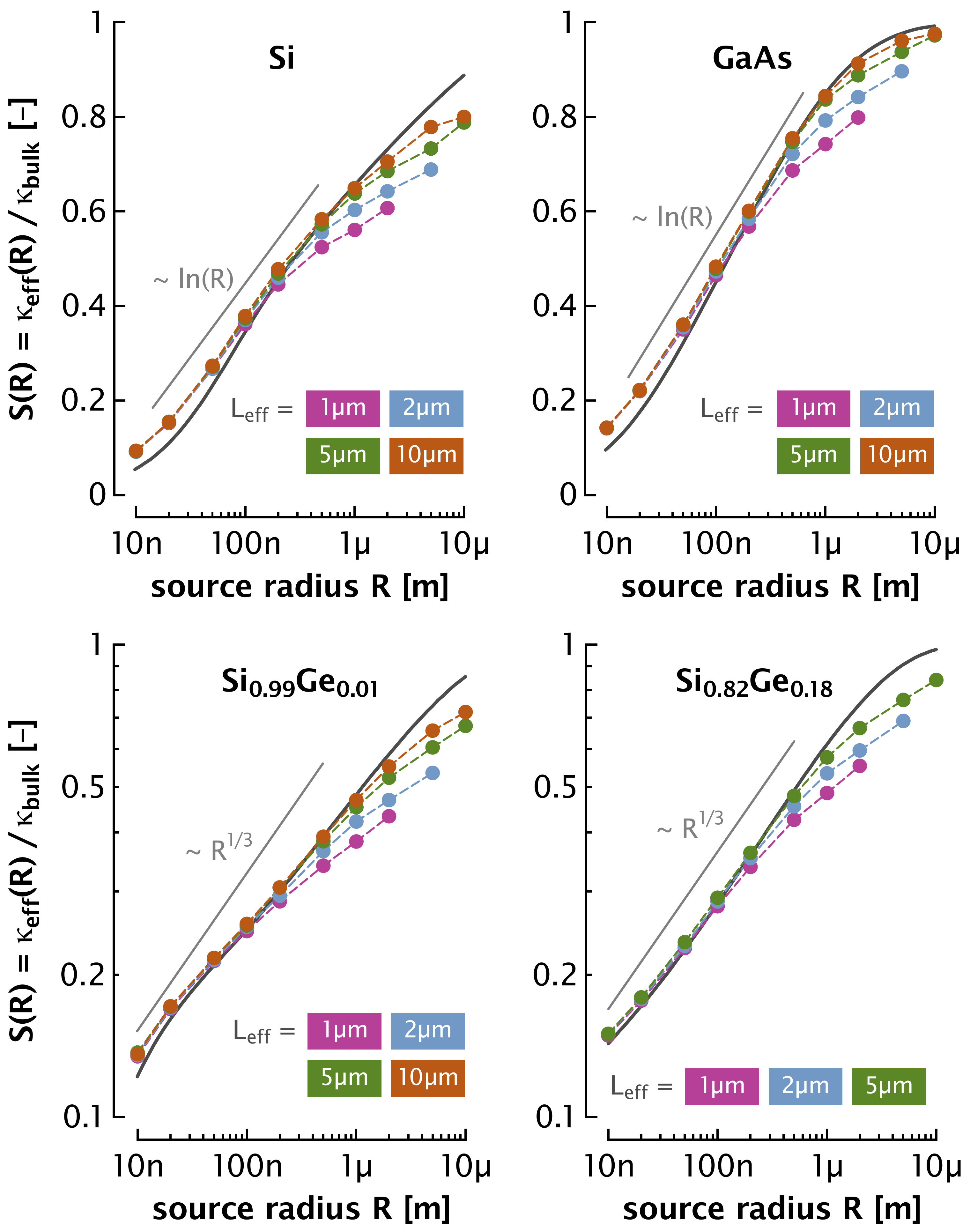}{Suppression function determined from first principles for spherical source configuration shown in Fig. \ref{fig1_sphericalgeometry}. Points connected by dashed lines obtained through Monte Carlo simulation; solid lines depict the semi-analytical infinite medium ($L_{\text{eff}} \rightarrow \infty$) solution per Eq. (\ref{Ssphere_numerical}).}
\par
The effective conductivity progressively decreases as the source gets smaller, dropping nearly an order of magnitude below the nominal bulk value for a 10$\,$nm source radius. This $R$-dependence is compounded by a secondary constriction induced by the finite simulation geometry (`multidimensional thin film' effect) that likewise gets worse for smaller $L_{\text{eff}}$. However, geometrically small sources ($R/L_{\text{eff}} \ll 1$) still `see' an infinite medium and in this limit we accordingly observe the Monte Carlo results for different $L_{\text{eff}}$ to indeed coincide. Interestingly, the semi-analytic solutions for the infinite medium (solid lines) do not precisely follow the same small-source asymptotes but rather fall slightly below the VRMC results. This can be attributed to differences in how the two approaches implement the heat dissipation inside the source. On the one hand, analytic BTE single pulse responses follow an energy-centric viewpoint by distributing the source power among the phonon channels according to their relative heat capacity contribution $C_k/\sum C_k$. The VRMC simulations, on the other hand, adopt a flux-based perspective and utilise phonon selection probabilities $C_k \| \vec{v}_k\| / \sum C_k \| \vec{v}_k\|$ for launching deviational particles from the source. Small differences between analytic and Monte Carlo curves aside, both first-principles results display the prototypical logarithmic (single crystals) and fractional power (alloys) dependences on source radius derived in the previous section.
\subsubsection{Circular source}\label{sec:results:vrmc2d}
The considered geometry is shown in Fig. \ref{fig3_circulargeometry}. Due to symmetry this configuration induces a thermal field that is half that produced by a circular source of radius $R$ integrated atop a rear-cooled substrate with thickness $L$.
\myfig[!htb]{width=0.4\textwidth}{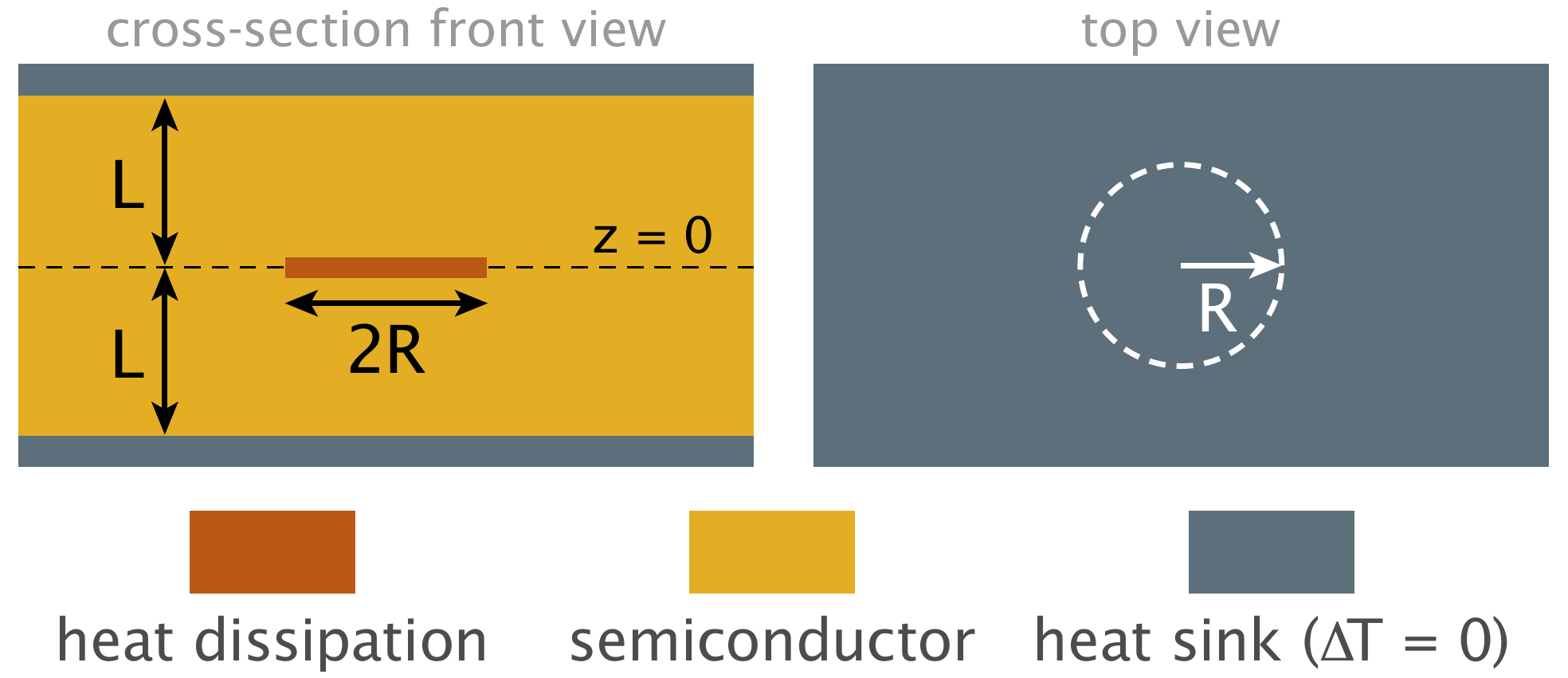}{Simulation geometry for circular heat source.}
\par
The diffusive response of the structure can be determined semi-analytically as follows. First, we derive the Hankel-domain Green's function of the infinite medium at planes parallel to the source:
\begin{equation}
G_{\text{Fourier}}(h,z) = \frac{1}{\pi} \, \int \limits_{0}^{\infty} \frac{\cos(\xi_z z) \, \mathrm{d}\xi_z}{D \left( h^2 + \xi_z^2 \right)} = \frac{\exp(-h |z|)}{2 D h}
\end{equation}
The heat sinks present in the finite slab structure can be accounted for by pairs of image sources (with alternating signs) in planes $z = \pm 2L, \pm 4L, \ldots$. The Green's function at the source plane in the finite slab evaluates to
\begin{multline}
G_{\text{Fourier}}^{\,\text{slab}}(h,z=0) = \frac{1}{2Dh} \times \\ \left[ 1 + 2 \sum \limits_{n=1}^{\infty} (-1)^n \, \exp(-h \cdot 2nL) \right] = \frac{\tanh(hL)}{2Dh}
\end{multline}
which subsequently yields for a circular source with uniform surface power density $p_s = Q_{\text{tot}}/\pi R^2$
\pagebreak[4]
\begin{equation}
\left< \Delta T_{\text{Fourier}} \right> = \frac{p_s R}{\kappa} \, \int \limits_{0}^{\infty} \frac{J_1^2(v)}{v^2} \, \tanh \left( v \, \frac{L}{R} \right) \, \mathrm{d}v
\end{equation}
In the geometrically thick limit $L/R \rightarrow \infty$ this correctly recovers the infinite-medium result $\left< \Delta T \right> = 4 p_s R/3 \pi \kappa$. For optimal numerical performance we once again integrate the oscillatory portion of the integrand analytically through use of the identity $\int [J_1^2(v)/v] \, \mathrm{d}v = \frac{1}{2} \left[ 1 - J_0^2(v) - J_1^2(v) \right]$. Comparison between the diffusive solution and VRMC counterpart results in the suppression curves shown in Fig. \ref{fig4_circular}.
\myfig[!htb]{width=0.48\textwidth}{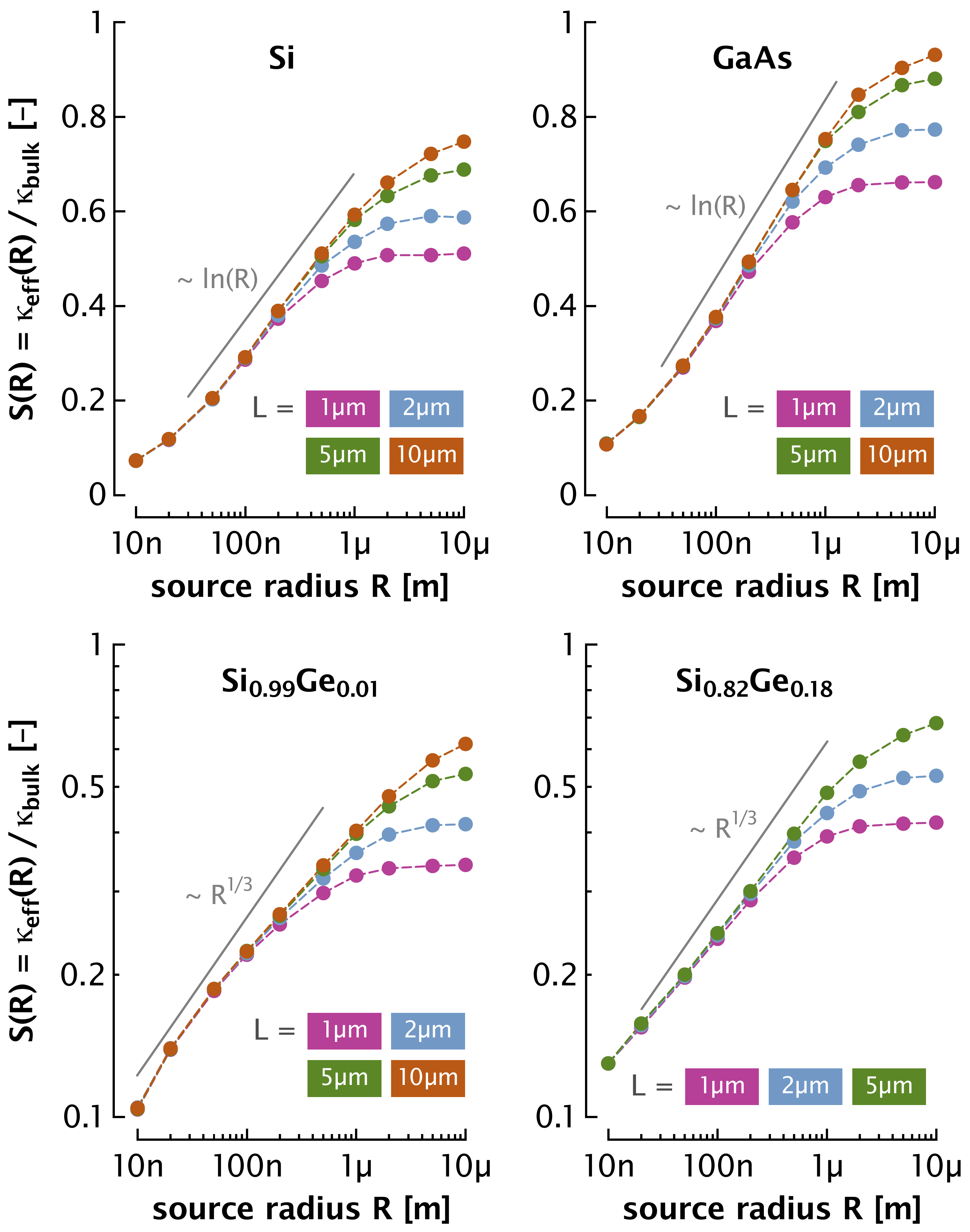}{Suppression function for circular source configuration illustrated in Fig. \ref{fig3_circulargeometry}. All results obtained by first-principles Monte Carlo simulation.}
\par
Similar to the spherical case discussed previously, a strong $R$-dependent constriction is compounded by a secondary thin-film suppression ($L_{\text{eff}}$ dependence) that disappears in the geometric `infinite' medium limit $R/L \ll 1$. Once again distinct quasiballistic signatures for single crystals and alloys are clearly visible.
\section{Additional discussion}\label{sec:discussion}
\subsection{Limitations of kinetic theory}\label{sec:discussion:kinetic}
It is interesting to note that the key signatures of the source-size dependent suppression, being ballistic and diffusive asymptotes $\kappa_{\text{eff}} \sim R$ and $\kappa_{\text{eff}} \simeq \kappa$ respectively and intermediate quasiballistic regimes $\kappa_{\text{eff}} \sim \ln(R)$ in single crystals and $\kappa_{\text{eff}} \sim R^{2-\alpha}$ in alloys, are identical to those previously observed \cite{APL-thinfilms} for the thickness dependent cross-plane conductivity $\kappa_{\perp}(L)$ in thin films. The question arises whether, as was the case for the thin films \cite{APL-thinfilms}, one may obtain accurate estimates simply by aggregating the suppression of individual phonon modes:
\begin{equation}
\kappa_{\text{eff}}(R) \stackrel{\text{?}}{\simeq} \sum \limits_{\omega} S_{\omega}^{\text{grey}}(\Lambda_{\omega},R) \cdot \kappa_{\omega} \label{kinetictheory}
\end{equation}
Viewed from a kinetic theory perspective the validity of (\ref{kinetictheory}) might be easily taken for granted, especially given that the studied problem is steady state and thus completely devoid of high-speed nonequilibrium effects. However, as we will illustrate for a spherical source, the approximation (\ref{kinetictheory}) tends to substantially underestimate the severity of the quasiballistic heat constriction.
\par
To derive $S_{\omega}^{\text{grey}}$ we start from the known \cite{minnich-BTE3D} spectral grey-medium single pulse response of the RTA-BTE:
\begin{equation}
P_{\omega}^{\text{grey}} = \frac{1}{\psi_{\omega}^{\text{grey}}(\zeta)} = \frac{\tau_{\omega} \arctan(K_{\omega} u)}{K_{\omega} u - \arctan(K_{\omega} u)} \label{Pgrey}
\end{equation}
where $K_{\omega} = \Lambda_{\omega}/R$ is the Knudsen number and $u = \zeta R$. We can now proceed exactly as in section \ref{sec:results:analytic} to obtain the suppression function through Eq. (\ref{Ssphere_numerical}). The diffusive and ballistic asymptotes can be derived in closed form: $S_{\omega}^{\text{grey}}(K_{\omega} \ll 1) \simeq 1/(1 + 2 K_{\omega}^2/3)$ and $S_{\omega}^{\text{grey}}(K_{\omega} \gg 1) \simeq 8/5 K_{\omega}$. The full curve (Fig. \ref{fig5_Sgrey}) including the quasiballistic regime can be accurately approximated (deviations remain within 3\%) by
\begin{equation}
S_{\omega}^{\text{grey}}(K_{\omega}) \simeq \left[ 1 + \left( 5 K_{\omega}/8 \right)^{1.35} \right]^{-\frac{1}{1.35}} \label{Sgrey_fit}
\end{equation}
\vspace{-3ex}
\myfig[!htb]{width=0.48\textwidth}{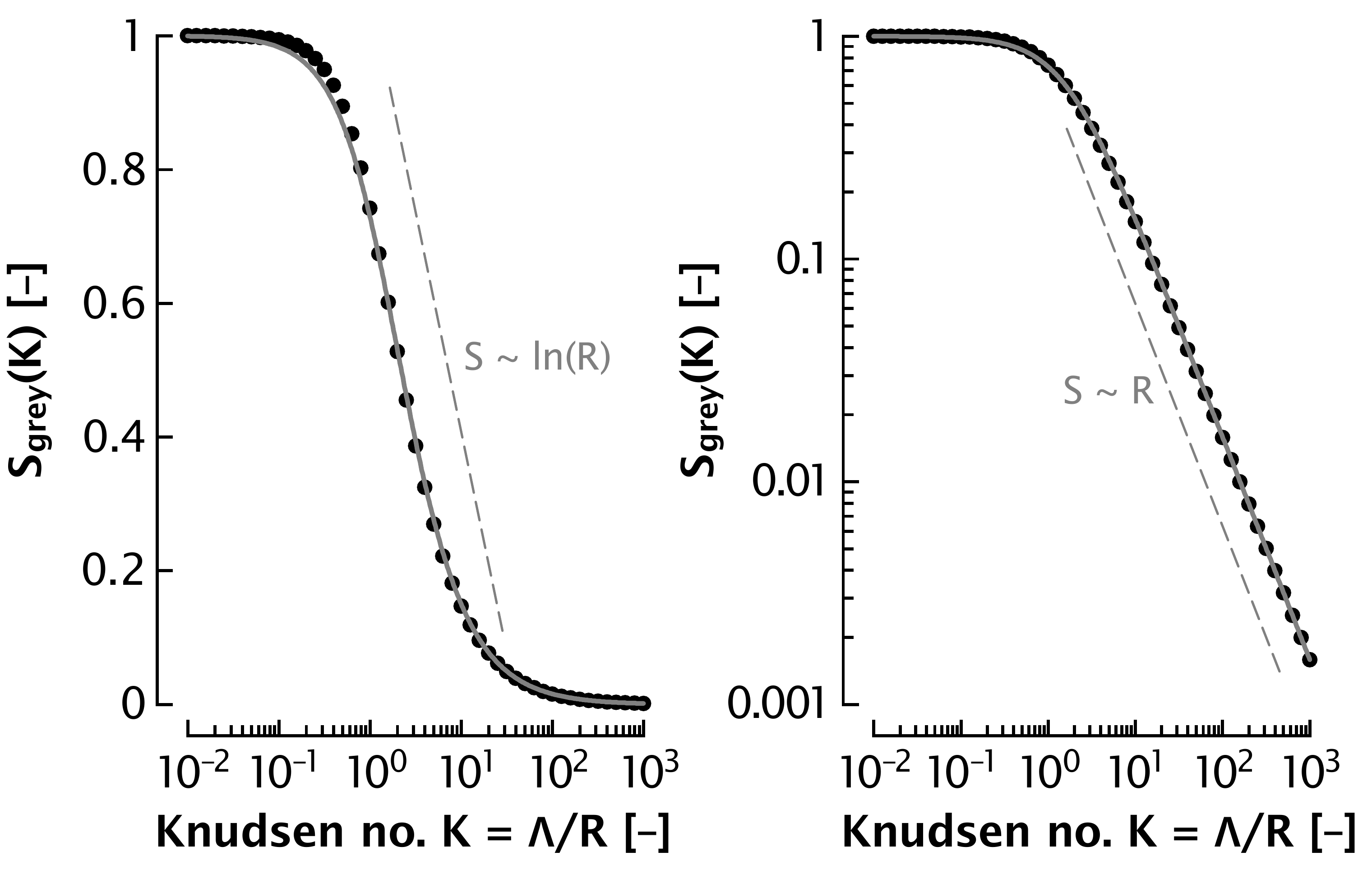}{Grey-medium suppression function for spherical source. The exact result (points) can be fitted by the compact functional form of Eq. \ref{Sgrey_fit} (solid line) within 3\%.}
\par
Comparing the kinetic estimates (\ref{kinetictheory}) to the previously discussed semi-analytic BTE solutions reveals that the former capture the main trends, but systematically overestimate the effective conductivity for small sources by up to threefold in the investigated single crystals and over twofold for the considered alloys (Fig. \ref{fig6_kinetic}).
\myfig[!htb]{width=0.48\textwidth}{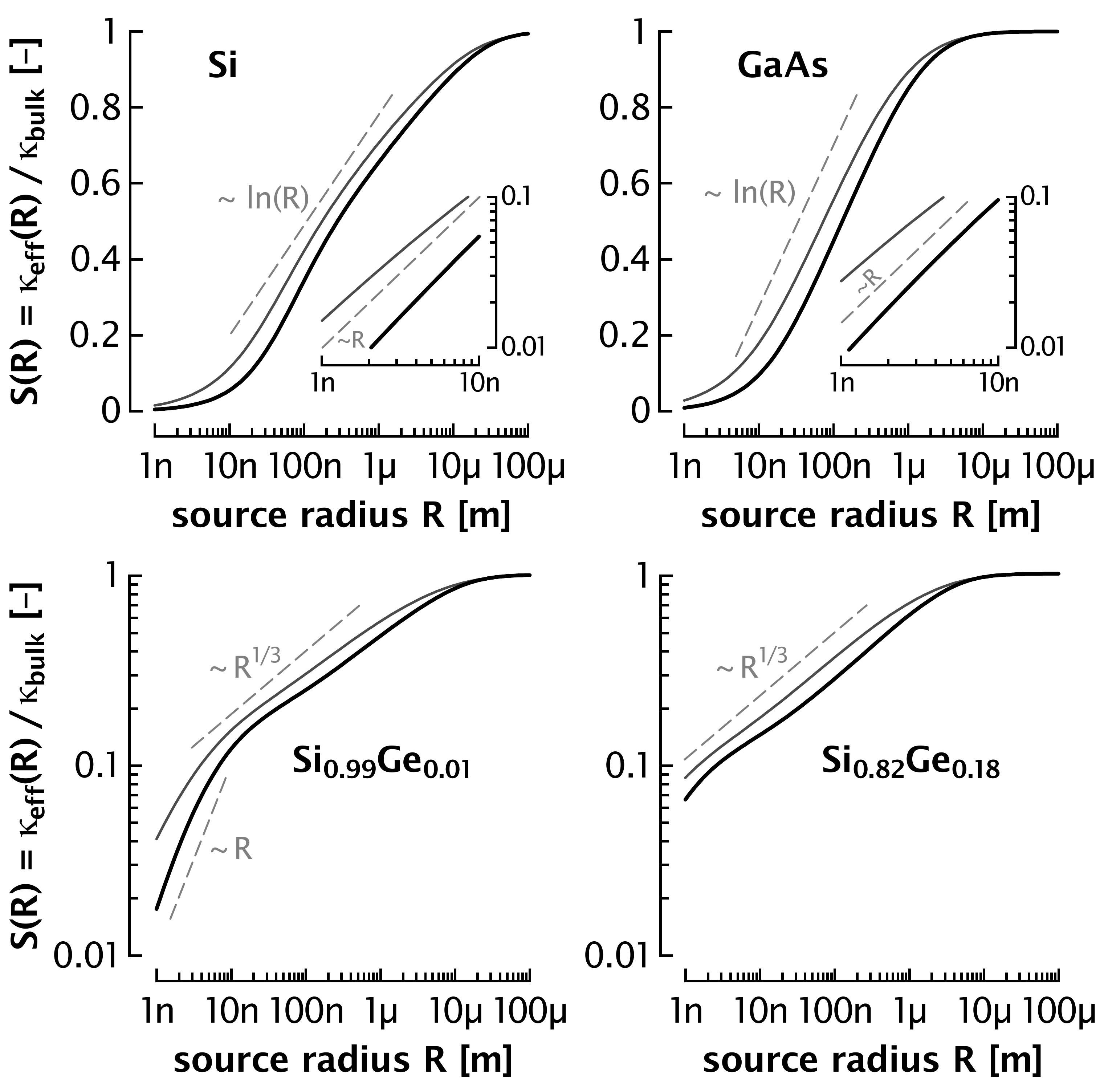}{Kinetic theory (aggregated grey-medium suppression, grey lines) systematically overestimates actual BTE solutions (black lines). All computations performed from first principles for spherical source in infinite medium. The double-logarithmic insets reveal the ballistic asymptotes.}
\subsection{Limitations of `modified Fourier' theory}\label{sec:discussion:modifiedfourier}
The size-dependent suppression function $S(R)$ offers an easy-to-use metric to quantify the deviation between the quasiballistic average temperature rise inside the source and its diffusive prediction. The seeming simplicity of this `modified Fourier' perspective may tempt device engineers to simply replace the bulk conductivity by its lower `effective' counterpart in hopes of obtaining more accurate thermal simulations. In contrast to successful applications in 1D thin film transport \cite{maassen}, however, adjusted diffusion theory is unable to properly capture the physics of the multidimensional heat conduction discussed here, and therefore remains confined to the phenomenological realm. Concretely, effective Fourier solutions for the spherical shell geometry [Eq. (\ref{Fourier_sphericalfield}) with $\kappa$ replaced by $\kappa_{\text{eff}}(R)$] display a weaker spatial decay and notably overestimate the thermal field in the vicinity of small heat sources (Fig. \ref{fig7_modifiedfourier}). We note that the Monte Carlo simulations for this particular analysis were performed with 100 times more particles ($N=10^7$), yielding highly reproducible and noise-free temperature profiles. (Unavoidable fluctuations very close to the orgin still remain, but these carry virtually zero weight in the spatial averaging through which $\kappa_{\text{eff}}$ is determined.)
\myfig[!htb]{width=0.48\textwidth}{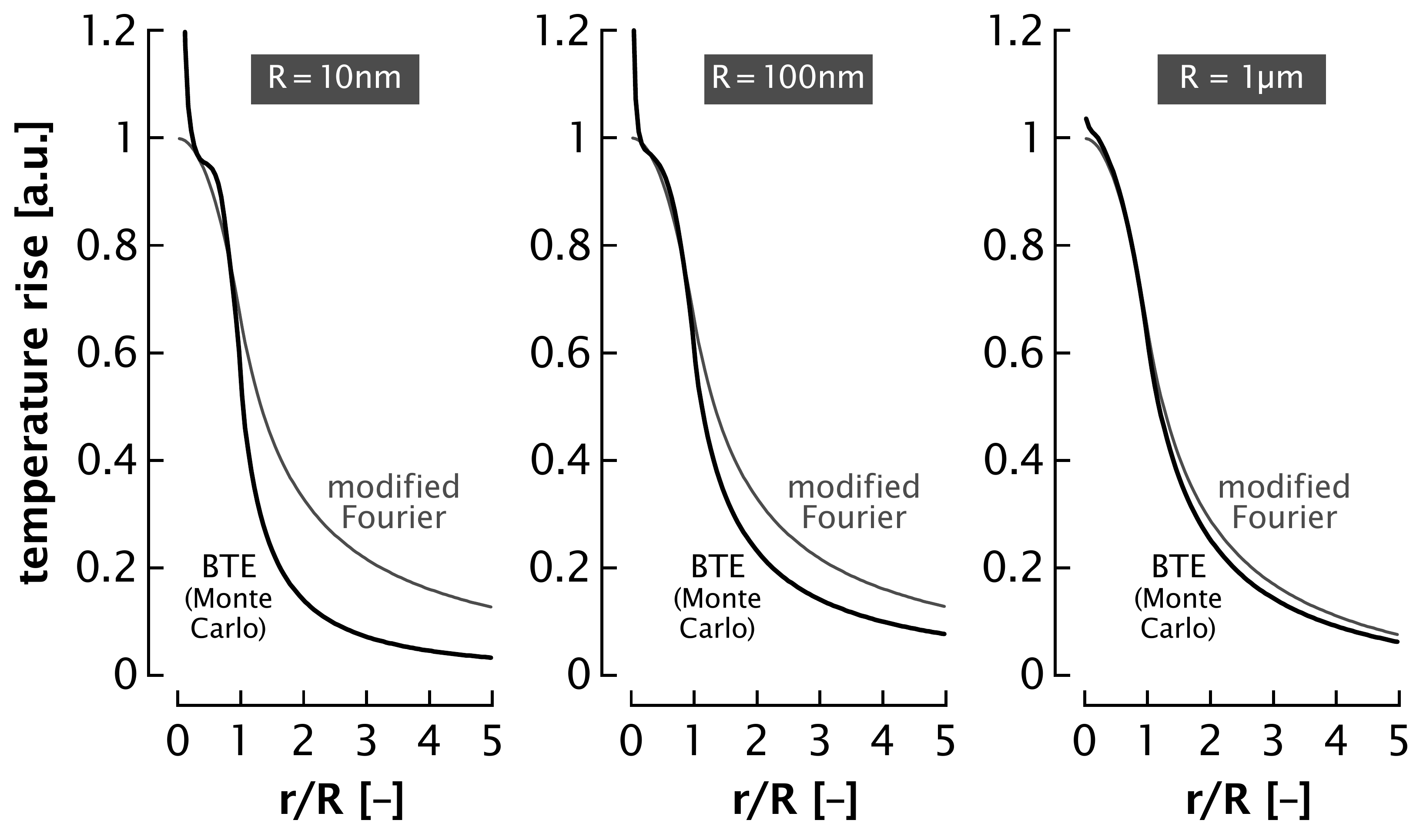}{Dimensionless thermal field in spherical Si configuration with $L_{\text{eff}} = 10\,\mu$m. Modified Fourier curves (diffusive solutions with effective conductivity so as to match the average quasiballistic temperature rise \textit{inside} the source) are a poor approximation to the BTE solution \textit{outside} the source.}
\subsection{`Ultraviolet' BTE divergence}\label{sec:discussion:uvdivergence}
As we touched upon earlier, attempting to analyse a circular source in a prototypical single crystal (\ref{psi_singlecrystal}) produced the nonsensical result of systematic thermal runaway ($\Delta T \rightarrow \infty$ irrespective of the source radius). The problem arises in the very first step of the calculation, where a divergence $G_0(h) \rightarrow \infty$ is observed for the Green's function in the source plane. The issue is rooted in the ballistic asymptote $\psi(\zeta r_0 \gg 1) \sim \zeta \leftrightarrow P(\| \vec{r} \| \ll r_0) \sim 1/\| \vec{r} \|^2$ of the single pulse response:
\begin{eqnarray}
P(r) & = & \lim \limits_{\lambda \rightarrow 0} \, \frac{1}{2 \pi^2 D} \, \int \limits_{0}^{\infty} \exp( -\lambda \zeta) \, (1 + r_0 \zeta) \, j_0(\zeta r) \, \mathrm{d}\zeta \nonumber \\
& = & \frac{1}{4 \pi D r} \, \left[ 1 + \frac{2 r_0}{\pi r} \right]
\end{eqnarray}
In the source plane $z=0$ this leads to $P(\rho \ll r_0) \sim 1/\rho^2$ which is not Hankel transformable [$J_0(h \rho)/\rho$ is not integrable near the origin, explaining the divergence of $G_0(h)$]. Even when operating within real space, the final result becomes infinite again: the central temperature rise of the circular source involves 2D integration of $P(\rho)$, which diverges logarithmically at the origin.
\par
While we used their prototypical propagator for illustration, it should be emphasised that the ballistic divergence is not limited to single crystals alone. The same asymptotic behaviour emerges also in alloy compounds (as stated before) and even in grey-medium solutions [Eq. (\ref{Pgrey}) readily shows that $\psi_{\omega}^{\text{grey}}(\zeta \Lambda_{\omega} \gg 1) \sim \zeta$].
\par
The integral anomalies encountered above are reminiscent of those previously observed in other physics disciplines such as quantum field theory \cite{quantumfield}. Commonly referred to as ultraviolet divergences (after classical mechanics famously predicting infinite blackbody radiation at high photon energies), these are indicative of phenomena taking place within the model framework at unphysically small distances \cite{uvdivergences}. In our context of phononic thermal transport, we can argue that here too a similar breakdown is at play. Indeed, the BTE and the Fourier transformations applied to solve it implicitly assume space to be infinitely divisible, whereas in reality heat is being transported through collective vibrations of discrete atoms around particular equilibrium positions. At length scales shorter than the lattice constant, thermal gradients and even the definition of temperature itself quickly lose their physical meaning.
\par
It remains to be seen whether the BTE could be `renormalised' to make it `UV complete'. In the meantime, Monte Carlo approaches such as those adopted here help to organically overcome the mathematical anomalies wreaking havoc in the semi-analytic solutions because these simulations sample the thermal field over a discrete mesh with finite spatial resolution.
%\pagebreak[4]
\section{Conclusions}\label{sec:conclusions}
In summary, whereas previous studies provided individual explanations for particular types of experiments, here we derived universal functional forms for the nondiffusive thermal constriction around small heat sources, and illustrated them with concrete first-principles analytic solutions and Monte Carlo simulations. The prototypical dependences of $\kappa_{\text{eff}}(R)$ on characteristic source dimension are found to be precisely the same as those previously established for cross-plane conductivity $\kappa_{\perp}(L)$ on film thickness, a fact which was not obvious at first sight. Even so, important distinctions between the two configurations arise, highlighting the physical intricacies of the quasiballistic heat conduction. In particular, kinetic and modified Fourier theories, known to yield accurate estimates for 1D thin film transport, perform quite poorly for the multidimensional problem studied here, and deviate by up to threefold from the actual BTE solutions. These findings can provide helpful insights for thermal design and analysis of nanoscale devices.
\section*{Acknowledgements}
The authors acknowledge funding from the \textsc{alma} Horizon 2020 project (European Union Grant No. 645776).
%\bibliography{biblio}% Produces the bibliography via BibTeX.
%INSERTED BBL FILE
\end{document}